\documentclass[conference]{IEEEtran}
\IEEEoverridecommandlockouts

\usepackage{cite}
\usepackage{xspace}
\usepackage{tikz}
\usepackage{needspace}
\usepackage{alltt}
\usepackage{amsfonts}
\usepackage{url}
\usepackage{hyperref}
\newcommand{\macaption}[2]{{\vspace{-10pt}}\caption{#2}\vspace{-8pt}\label{#1}}
\newcommand{\omg}{$\Omega$\xspace}
\newcommand{\lisp}{\texttt{Lisp}{}\xspace}

\newcommand{\eiffel}{\texttt{Eiffel}{}\xspace}
\newcommand{\self}{\texttt{S{\sc{}elf}}{}\xspace}
\newcommand{\smalltalk}{\texttt{Smalltalk}{}\xspace}

\newcommand{\elit}{{\it{}\raisebox{1px}{e}}\hspace*{-1pt}{$\mathcal{L}\mathfrak{i}$}{\it{}t}\xspace}

\newcommand{\java}{\texttt{Java}{}\xspace}

\newcommand{\prograph}{\texttt{Prograph}{}\xspace}
\newcommand{\logo}{\texttt{Logo}{}\xspace}
\newcommand{\scratch}{\texttt{Scratch}{}\xspace}
\newcommand{\snap}{{\it{}Snap!}{}\xspace}

\newcommand{\lldots}{.\hspace{-0.2pt}.\hspace{-0.2pt}. }
\newcommand{\git}{{\it{}Git}{}\xspace}
\newcommand{\svn}{{\it{}Subversion}{}\xspace}
\newcommand{\gitsvn}{{\git}/{\svn}{}\xspace}

\definecolor{commentcol}{RGB}{159,0,10}
\newcommand{\comm}[1]{\textcolor{commentcol}{#1}}

\definecolor{methcol}{RGB}{0,0,255}
\newcommand{\meth}[1]{\textcolor{methcol}{#1}}

\definecolor{localcol}{RGB}{0,0,0}
\newcommand{\local}[1]{\textcolor{localcol}{#1}}

\definecolor{TYPECOL}{RGB}{0,118,17}
\definecolor{typecol}{RGB}{0,118,17}
\newcommand{\type}[1]{\textcolor{typecol}{#1}}
\newcommand{\pargen}[1]{%
\textcolor{typecol}{\tikz[baseline=(char.base)]{%
      \node[shape=rectangle,draw,rounded corners=0pt,line width=1pt,inner sep=1.8pt] (char) {\small{{\tt{}#1}}};}}%
}

\newcommand{\typegen}[2]{\type{#1}\,\pargen{#2}}

\definecolor{opercol}{RGB}{145,76,38} %153,21,75}
\newcommand{\liop}[1]{\textcolor{opercol}{#1}}
\newcommand{\andand}{\textcolor{opercol}{\tt\&\&}}

\newcommand{\add}{\textcolor{opercol}{\bf\tt{}+}}

\newcommand{\ambigfig}{{\includegraphics[scale=0.1]{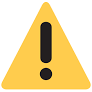}}}

\definecolor{rececol}{RGB}{128,128,128}
\definecolor{selfcol}{RGB}{145,76,38}
\newcommand{\selfrece}{\textcolor{rececol}{{\tikz[baseline=-3pt]{\node[shape=circle,line width=1.5pt,draw=gray,inner sep=1.5pt] (char) {\,};}}}}
\newcommand{\selftype}{\textcolor{typecol}{{\tikz \draw[line width=1.3] (0,0) circle (0.11cm);}}}
\newcommand{\selfself}{\textcolor{selfcol}{{\tikz \draw[fill=selfcol] (0,0) circle (0.09cm);}}}
\newcommand{\selfclone}{\textcolor{rececol}{{\tikz[baseline=-3pt]{\node[shape=circle,line width=1.5pt,draw=gray,inner sep=1.5pt] (char) {\,};}%
    \hspace{-0.4em}\tikz[baseline=-3pt]{%
      \node[shape=circle,line width=0.5pt,draw=gray,inner sep=1.5pt] (char) {\,};}}}}
    
\newcommand{\assignarrow}{{\tikz {\draw[line width=0.9pt,draw=black]
    (2.5pt,2.5pt) -- (0pt,0pt) -- (2.5pt,-2.5pt);
    \draw[line width=0.9pt,draw=black] (0.01pt,0pt) -- (10pt,0pt);}}}

\newcommand{\typz}{\textcolor{typecol}{$\mathbb{Z}$}}
\newcommand{\typnn}[1]{\textcolor{typecol}{$\mathbb{N}_{#1}$}}
\newcommand{\typnnexp}[1]{\textcolor{typecol}{$\mathbb{N}^{#1}$}}
\newcommand{\typnnplain}[1]{\textcolor{typecol}{$\mathbb{N}{#1}$}}
\newcommand{\typzn}[1]{\textcolor{typecol}{$\mathbb{Z}_{#1}$}}

\newcommand{\typrn}[1]{\textcolor{typecol}{$\mathbb{R}_{#1}$}}

\definecolor{textcol}{RGB}{254,115,33}
\definecolor{backdark}{RGB}{1,37,0}

\definecolor{colOperator} {RGB}{145, 76, 38}
\newcommand{\ka}[2]{
  \IfSubStr{#1}{R}{$\tikz[baseline=-0.2]{%
    \draw [<<-,line width=1pt,draw=colOperator] (0,3pt) -- (15pt,3pt);
    }^{\hspace{-6pt}\textbf{\textcolor{colOperator}{\tiny{#2}}}}$%
  }{%
    $\tikz[baseline=-0.2]{%
    \draw [->>,line width=1pt,draw=colOperator] (0,3pt) -- (15pt,3pt);
    }^{\hspace{-14pt}\textbf{\textcolor{colOperator}{\tiny{#2}}}}$%
  } %
}

\newcommand{\mafigpng}[2]{%
\begin{figure}[ht]%
\vspace{-5pt}\centering%
\includegraphics[width=1.0\columnwidth]{#1.png}%
\macaption{#1}{#2}%
\end{figure}%
}

\newcommand{\mafignimp}[4]{%
\begin{figure}[ht]\centering%
\includegraphics[width=#1\columnwidth]{#3.#2}%
\macaption{#3}{#4}%
\end{figure}%
}

\newcommand{\utf}[1]{{\sc{utf}}{\tt{}-}{\small\tt{}#1}{}\xspace}

\begin{document}
\setlength{\tabcolsep}{2pt}

\title{{\omg}: The Power of Visual Simplicity}

\author{\IEEEauthorblockN{Benoît Sonntag}
\IEEEauthorblockA{\textit{Université de Strasbourg}, Strasbourg, France \\
Benoit.Sonntag@lisaac.org - 0009-0000-2806-1970}
\and
\IEEEauthorblockN{Dominique Colnet}
\IEEEauthorblockA{\textit{Université de Lorraine - LORIA}, Nancy, France \\
Dominique.Colnet@loria.fr - 0009-0006-7368-2235}}

\maketitle

\begin{abstract}%% max 250 words
We are currently developing an innovative and visually-driven programming language called {\omg} (Omega).
Although the {\omg} code is stored in text files, these files are not intended for manual editing or traditional printing.
Furthermore, parsing these files using a context-free grammar is not possible.
The parsing of the code and the facilitation of user-friendly manual editing both necessitate a global knowledge of the codebase.
Strictly speaking, code visualization is not an integral part of the {\omg} language; instead, this task is delegated to the editing tools.
Thanks to the global knowledge of the code, the editing process becomes remarkably straightforward, with numerous automatic completion features that enhance usability.
{\omg} leverages a visual-oriented approach to encompass all fundamental aspects of software engineering.
It offers robust features, including safe static typing, design by contracts, rules for accessing slots, operator definitions, and more,
all presented in an intuitively and visually comprehensible manner, eliminating the need for obscure syntax.
\end{abstract}


\section{Introduction}
The thirty glorious years of computer languages from 1960 to 1990 gave rise to most of the major concepts of today's programming.
This flourishing period has touched every major issue in computer programming.
First of all, memory management, with the notion of execution stack and the appearance of the first garbage collector ({\lisp} \cite{lisp1,lisp2}).
During the same period, different directions concerning code management have emerged, with the fusion of code and data (again {\lisp}),
or on the contrary a more formal approach with functional languages
(ML \cite{ml}, Caml \cite{caml}, Ocaml \cite{ocaml}).
Then a more pragmatic approach, with a grouping of code and data present in object languages (Simula 67 \cite{simula67}, {\smalltalk} \cite{smalltalk80})
and the notion of inheritance which reaches its apogee with the prototype model ({\self} \cite{ungar87b}) and its dynamic inheritance.
In this context, the semantics of languages was the battle horse to the detriment of form, i.e. notation and syntax, which took a back seat.

For historical reasons that we can well understand, the almost systematic use of context-free grammars, of either kind $LL$ or kind $LR$, has never been really questioned.
We can only observe the bogging down of textual syntaxes and the appearance of real religious wars concerning syntactic choices.
%%For this question, the computer language industry has decided: only C-oriented notations remain and the time for innovation or originality is over.
We believe that traditional context-free grammar has reached an evolutionary dead end and does not benefit from technological advances.
Today's computers or phones offers us much more sophisticated representation possibilities.
Also, we see appearing little by little an approach richer of communication between the man and the machine than the simple keyboard\,/\,mouse.
It is still difficult today to predict the impact and interest of the appearance of multiple sensors, tactile, visual, auditory, \lldots
that we already have in our phones or computers.
But, it seems obvious to us that there is a growing gap between the hardware and the programming environments we use.

Furthemore, in a way too often decorrelated to expressiveness, for example for the optimization of $GPU$, 
compilers have evolved in terms of performance, but mostly for low-level languages.
This almost immutable image in the collective mind of programmers, of the impossibility of combining performance and expressiveness must end.
Let's note however some efforts to marry performance and high level language
like the OCaml compiler \cite{ocaml}, the SmartEiffel compiler \cite{bdboopsla97}, or more recently with Rust\,\cite{rust} and
Lisaac\,\cite{sonntag2002a,sonntag2002b}.

We aim to continue these efforts and to break with these various limitations and prejudices.
{\omg} is intended to be a high-level language with a graphic visual while maintaining a rigorous terminology.
Here we are talking about denotational terminology related to the field of application.
The goal is to get rid of a rigid syntax to adopt the usual notations of a domain, as for example mathematics, physics, etc.

\section{The {\omg} language at a glance using ELIT eyes}
\subsection{Visualized Code Blocks and Auto-Indentation}
The most significant innovation in the {\omg} language is the concept of a visual approach that opens up a wider range of possibilities and notations.
In contrast to its predecessor, the Lisaac language, a {\omg} program is not constructed as a text file bound by established and rigid syntax.
Instead, a {\omg} program is represented as an editable figure using a specialized tool called {\elit},
which allows you to visualize and edit the figures and pieces of text that comprise a program.

\mafigpng{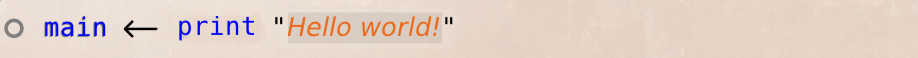}{
  The usual {\it{}Hello world!} program.
}
Figure \ref{elit-hello} shows an image capture inside our {\elit} development environment.
This is the traditional {\it{}Hello World!} program in {\omg}.
All the code visualization and, in particular, all the colors used for rendering are not part of the language definition.
The choices for displaying and editing programs can be redefined and are entirely up to the editing tool.
Note that {\elit} is an integrated development environment for {\omg}, itself written in
{\omg}\footnote{As most video games, {\elit} relies directly on the GPU thanks to the {\omg} library.
The same {\omg} code runs on most platforms\,/\,OSes.
Following link is a demo of {\elit}: \url{http://ks387606.kimsufi.com/omega}\\
The first release of {\omg} will be available in a few months.
}.
\mafigpng{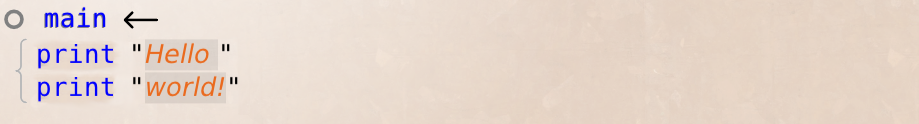}{
  Two \meth{print} statements within a multi-line block.
}
Figure \ref{elit-hello2} shows a variant of the previous program that uses two \meth{print} instructions to perform the same display.
The block that groups the two instructions is represented by an opening curly brace on the left.
Without any explicit intervention on the part of the programmer, {\elit} automatically selects the code representation.
Its choice of representation in a given context is based on statistics carried out on existing code.
In all cases, the display tool guarantees indentation and visualization in accordance with the code's semantics.
When two instructions are displayed on the same line, {\elit} may utilize a semicolon as a separator or employ other types of visual effects.
Nothing is frozen by language rules.
\mafigpng{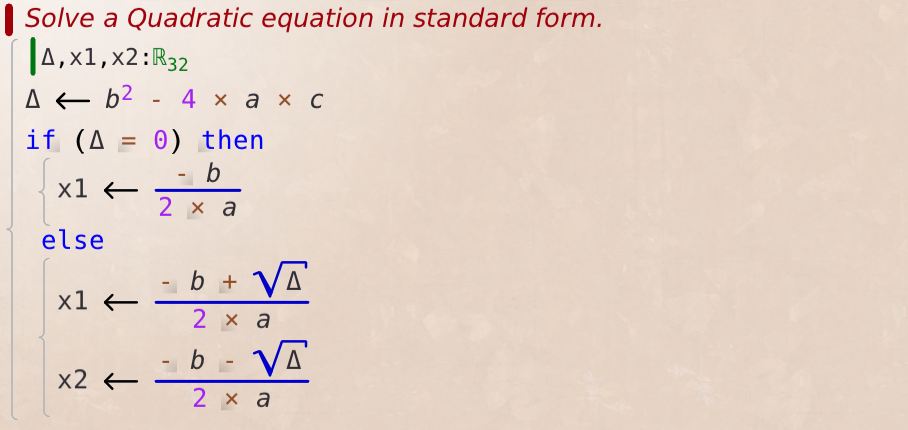}{
  The default is to visualize \comm{comments} in red and \type{types} using green.
}

\subsection{Bringing Math and Computer Notations Together}
The figure \ref{elit-maths} example starts with a comment, which can be identified by its red color, the default color of {\elit} for displaying comments.
The {\omg} language also makes it possible to add images, links and videos within comments.
The enclosing block comprises three local variables \local{$\Delta$}, \local{\tt{}x1} and \local{\tt{}x2}, all of which are of type {\typrn{32}}, a 32-bit floating-point number.
Note that variable or type names can be composed of any unicode characters (in \utf{8} format), with the added possibility of subscript or superscript formatting.
We've also resolved the debate over CamelCase {\it{}vs.} snake\_case conventions by allowing spaces within any type of identifier.

\mafigpng{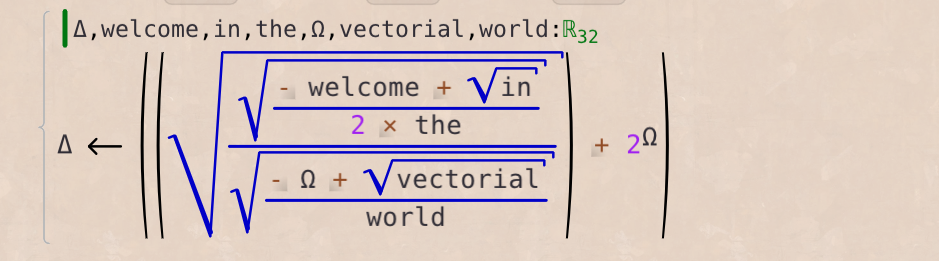}{
  Graphical symbols in blue color on all figures are vectorially defined.\\
  ~~
}
{\omg} is a statically typed language and all variables, attributes and arguments have a type.  
You'll also notice that {\elit} visually identifies types with a green color, as for example:
{\typrn{32}}, {\typz}, {\typzn{32}}, {\typzn{64}}, {\typnn{8}}, {\type{String}}, {\type{Boolean}}, etc.
To avoid confusion between comparisons and assignment, we use the algorithmic arrow \assignarrow{} for assignment.
Comparison operators are naturally those used in mathematics, such as
\liop{$=$}, \liop{$\ne$}, {\liop{$<$}},{\liop{$\le$}}, etc. 
For mathematics (still in figure \ref{elit-maths}), more advanced symbols already exist in the library:
fraction, square root, rise to power, sum, integral, etc.
These symbols are defined vectorially (see figure \ref{elit-bigsqrt}) using a technology inspired by Tikz from \LaTeX \cite{tikz2020}.

\subsection{Domain Specific Graphical Operators}
If a new symbol is needed for a specific domain, it can easily be defined and added to the library.
As an example, figure \ref{elit-electronic} shows the use of logic gates to create an adder.
\mafigpng{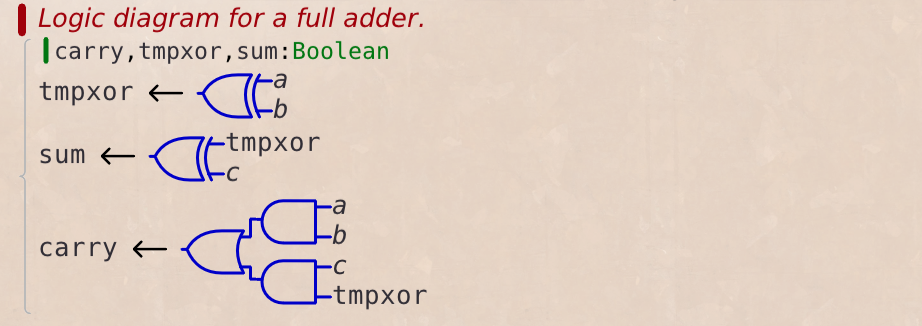}{
  Example of a user-defined electronic symbol.
}
In the following, we'll see how graphic symbol signatures integrate simply and seamlessly with more common textual signatures.
Once a new graphical operator is defined, it becomes available for use in any other prototype or context.
In addition to custom operators, {\omg} includes all the standard control structures such as \meth{for} loops,
\meth{foreach} loops,
\meth{while} loops,
\meth{until} loops,
\meth{switch} statements, and more.
It's worth noting that all these control structures, including the \meth{if\,then\,else} construct depicted in Figure \ref{elit-maths}, are implemented as library definitions.

\subsection{Extra Built-in Literal for Two Dimentional Notations}
In addition to literals for strings or numeric constants, {\omg} also has literals for arrays, tuples of types or sequences of values.
Thanks to its graphic and visual approach, {\omg} provides a natural way of visualizing a two-dimensional sequence of elements.
We felt it was essential to add a literal for a pleasant representation of matrices (see figure \ref{matrix}).
\mafigpng{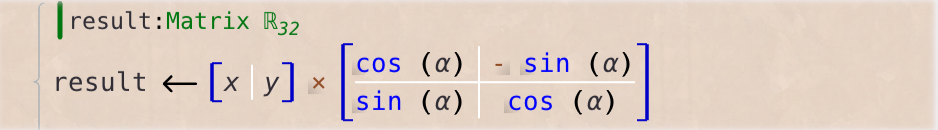}{
  The {\omg} language also incorporates a two-dimensional built-in literal.
}
Unlike all the other graphic operators that can be defined in libraries, this two-dimensional literal is a built-in part of the language.

\section{Signature for Slots}
As previously mentioned, the language is statically typed, and all operations, whether they are textual or graphical, have a precise
signature\footnote{As {\omg} is a prototype-based language, we prefer to use the term {\it{}signature} to provide information about the function parameters and return type.}.
Before we address how the signatures of graphical operators are specified, let's start with what is more common: textual signatures.
\mafigpng{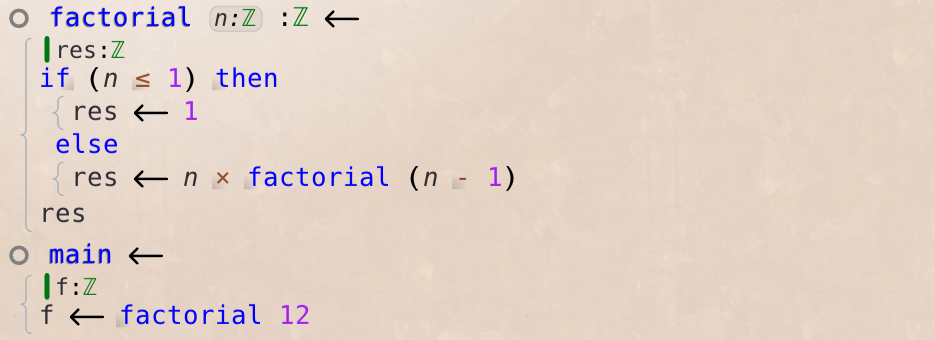}{
  Strong static and explicit typing everywhere.
}

\subsection{Textual Method or Attribute Signature}
In the example shown in figure \ref{elit-factorial},
we indicate both that the argument \local{{\it{}n}} of the \meth{factorial} function is of type {\typz} and that the result type of this function is also {\typz}.
The result returned by the \meth{factorial} function is the expression that ends the function block.
In this case, the result is the reading of the local variable \local{{\tt{}res}}, also of type {\typz}.
The \meth{main} procedure shows an example of how to call the \meth{factorial} function.
Note that with the chosen signature for the definition of \meth{factorial}, the use of parentheses is not required in the calling form.

\mafigpng{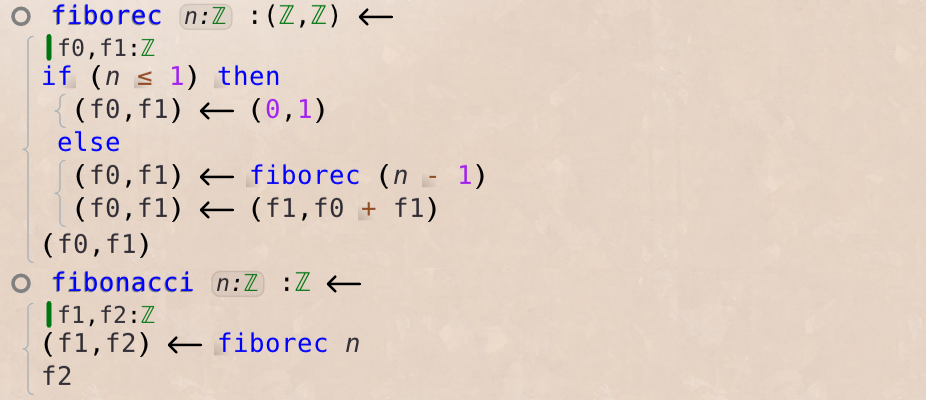}{
  The \meth{fiborec} function return a pair of {\typz} integers.
}
If required, it is also possible to define a tuple of types in the case of a function with several results. 
In the example of the figure \ref{elit-fibonacci}, the function \meth{fiborec} returns a pair of values, each of type {\typz}.
So the pair \local{{\tt{}(f0,f1)}} is a valid result for this function.
The \meth{fibonacci} function is implemented by a call to the recursive \meth{fiborec} function.
Note that this possibility of easily returning several results, without allocation in the heap, makes it possible here to obtain a recursive version in $O(n)$.
This notion of tuple is generalized to all type declarations.
In this way, a result tuple can be directly injected into an argument tuple.

\subsection{Graphical Operator Signature}
\mafigpng{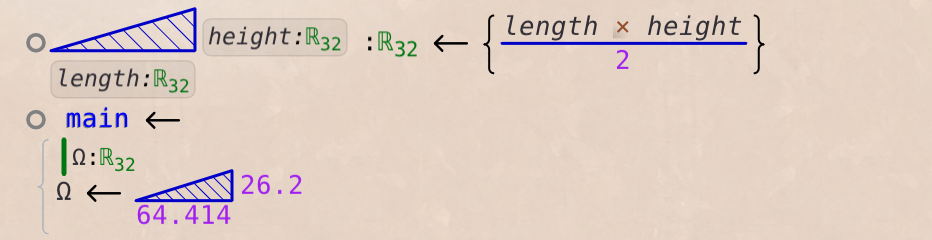}{
  A graphical operator with 2 parameters.
}
The visual display for defining graphical operators is identical to that used when using them.
Figure \ref{elit-triangle} shows an example of the signature for a triangle-shaped graphical operator.
This operator has two parameters, \local{{\it{}height}} and \local{{\it{}length}}, both of type \typrn{32}.
The result type of this operator is also \typrn{32}. 
In the example, the surface area of the triangle is the result.
The main procedure below provides an example of its usage.

As we will see in more detail in section \ref{visual}, each graphical operator is identified by a textual signature.
This signature serves both to facilitate code editing and to determine the order of evaluation.
For example, regarding the square root, the corresponding function name is simply {\tt{}sqrt}.

\section{Object Oriented Programming is Back}
{\omg} is a pure prototype-based programming language where every entity corresponds to a prototype.
This rule applies to all objects, from the most complex to the most basic.
For example, when it comes to numeric types, like signed integers, the {\omg} language allows you to use the {\typz} type to indicate a number of any size, with no limit other than the size of RAM memory.
As {\omg} also aims to program any type of device with the best possible performance, it is also possible to explicitly use {\typzn{8}}, {\typzn{16}}, {\typzn{32}} or {\typzn{64}}.
One can indicate that the memory representation of some prototype must be directly mapped onto some machine word.
Here, {\typzn{8}} is mapped on a 8-bit signed integer, {\typzn{16}} is mapped on 16-bit signed integer, and so on.
In case of direct mapping, there is no indirection, i.e. no pointer, and passing a {\typzn{32}} prototype is identical to passing an {\tt{}int32\_t} value in the C language.
The implementation of the general type {\typz} relies on attributes used to store arrays of elementary values, enabling the manipulation of large values.
Regardless of whether an object is elementary and mapped or defined by its attributes, it is necessary to be able to visualize or designate the receiver as well as its type.

\subsection{Graphical Visualization of the Receiver and its Type}
\mafigpng{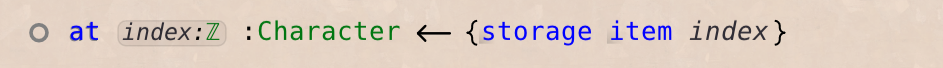}{
  Definition of method \meth{at} in \type{String}.
}
To simplify the presentation, in the preceding examples, we intentionally omitted specifying that the \selfrece{} disk indicates the receiver's position in the signature.
Any instruction or expression is always within a prototype, and this prototype is always passed implicitly if necessary.
Thus, there are no functions or procedures without a receiver.
For example, in the code from Figure \ref{elit-factorial}, all calls to the \meth{factorial} method pertain to the prototype in which this code is written.
In this case, as in all the others seen previously, the called method only uses the data provided by the arguments.
The current prototype is implicitely passed but not used.

Let's now explore the \meth{at} method as defined in the \type{String} prototype and illustrated in Figure \ref{elit-stringat}.
The purpose of this function is to access a character at a given index in the receiving string.
Outside the context of the String prototype, it is necessary to specify which string of characters you want to work with.
Therefore, the expression to access the character at index \local{i} in a character string \local{s} is simply written as:
\hspace{3mm}\local{s}\,\meth{at}\,\local{i}

The leftmost circle \selfrece{} indicates the position of the receiver just before the \meth{at} keyword within the signature.
Please be aware that using a period in the calling form is not required.
In this case, the calling form, (\local{s}\,\meth{at}\,\local{i}), is identical to the notation used in {\smalltalk} or {\self}.

\mafigpng{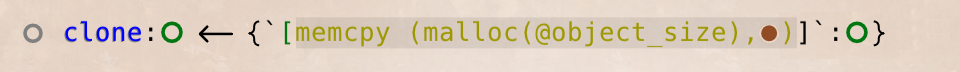}{
  The green circle indicates that the result matches the receiver type.
}
Within the body of a method, the \selfself{} symbol is used to designate the receiver of the message.
It's the equivalent of {\tt{}self} in {\smalltalk} or {\tt{}this} in {\java}.
The symbol \selftype{}, allows you to note the exact type of receiver using the idea presented in \cite{colnet2000b}.
This choice ensures safe typing while at the same time avoiding the criticisms made to {\eiffel} on this subject\cite{colnet2000b}.
Figure \ref{elit-clone} presents the signature of the \meth{clone} method for which the exact type of the receiver is particularly well-suited.

\subsection{Multiple Inheritance, Polymorphism and Visual Selectors}
Inheritance in {\omg} is multiple, and each inheritance link indicates whether polymorphism is authorized or not.
Furthermore, to facilitate the understanding of inheritance choices, one can decide whether to visualize inherited methods or not.
Of course, in {\omg}, all these choices are made
graphically.
First example of Figure \ref{inherit-boolean} presents the inheritance tree of the \type{Boolean} prototype.
A dotted line indicates implementation inheritance: all slots are inherited, but without permission for assignment.
All descendant prototypes are also displayed.

\begin{figure}[htbp]
  \centering
  \begin{minipage}{.5\columnwidth}
    \centering
    \includegraphics[width=0.95\linewidth]{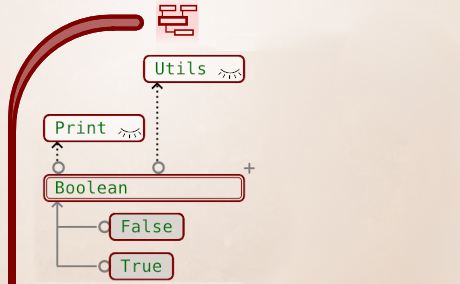}
    \macaption{inherit-boolean}{No syntax for inheritance. The graphic is directly editable.}
  \end{minipage}%
  \begin{minipage}{.5\columnwidth}
    \centering
    \includegraphics[width=0.95\linewidth]{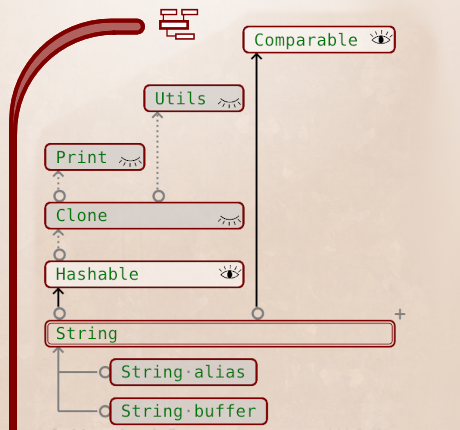}
    \macaption{inherit-string}{Open eyes are used solely to assist users in browsing the code.}
  \end{minipage}
\end{figure}

So in the example of \type{Boolean} we see that there are two descendant prototypes, \type{False} and \type{True}, each of which can be the object of an assignment in \type{Boolean}.
Type \type{Boolean} inherits \type{Print} and then \type{Utils}, each time with a dotted line.
These two prototypes, \type{Print} and \type{Utils}, are in fact simple method tanks \cite{gof94}.
Note that this inheritance visualization made by {\elit} is also the one used to navigate or to modify inheritance.
Once again, thanks to the visual approach, there's no need to learn any special syntax.
Unlike a standard textual approach, we have a global representation of the inheritance tree, including the descendants.

Let's take a look at the \type{String} inheritance graph shown in figure \ref{inherit-string}. 
According to the selected inheritance order, \type{Hashable} before \type{Comparable}, the visualization shows the lookup algorithm's scanning order from bottom to top.
Thus, if the lookup of some slot starts from \type{String}, the order of search is as follows:
\type{Hashable} {\small\bf\tt{}(1)},
\type{Clone} {\small\bf\tt{}(2)},
\type{Print} {\small\bf\tt{}(3)},
\type{Utils} {\small\bf\tt{}(4)},
\type{Comparable} {\small\bf\tt{}(5)}.
If you need to modify the order of inheritance or the nature of an inheritance link (solid or dotted line), you can do so graphically via {\elit} or by using keyboard shortcuts.
It's essential that the editing tool is sufficiently easy and intuitive to use.
To achieve this, we drew inspiration from video games as well as GPU capabilities.

In the {\omg} prototype library, there are two types of \type{String}:
immutable and aliased strings of type \type{String alias},
and extensible and modifiable strings of type \type{String buffer}.
Of course, all \utf{8} characters can be used with natural indexing.
One of the problems of object-oriented programming is the lack of visibility of inherited slots when going through the code of a prototype.
In Figure \ref{inherit-string}, the open eye on \type{Hashable} and \type{Comparable} allows you to view their inherited slots directly within the \type{String} edition.
A closed or open eye does not affect the semantic of the code.
This is purely a visual artefact for the convenience of the code designer.
Thus still on the example in figure \ref{inherit-string}, the eyes is close on \type{Clone} and \type{Print} because those prototypes are very familiar.
As we are well aware of the inherited slots from \type{Clone} and \type{Print}, we thereby prevent their visualization in the child prototypes.

To further discuss {\omg} inheritance, it's important to note that there is no default root for the inheritance tree.
It is entirely possible to create a new prototype that inherits nothing, a practice often beneficial, especially when modeling singletons \cite{gof94}.
In order for a prototype to be cloneable, it must inherit, either directly or indirectly, from the \type{Clone} prototype, which contains the \meth{clone} method
– the sole method responsible for generating an object by duplicating an existing prototype.
As \type{True} and \type{False} prototypes do not inherit from \type{Clone}, they remain unique objects.

\subsection{More Flexibility to Place the Receiver in the Signature}
As the {\omg} language is above all perfectly object-oriented, it's thanks to dynamic binding that conditional instructions are implemented.
Figure \ref{elit-boolean} shows an extract from the code of \type{Boolean}, which corresponds to the definition of the \meth{if then else} construct used in the previous examples.
The small round \selfrece{} in the signature indicates the receiver's position.
\mafigpng{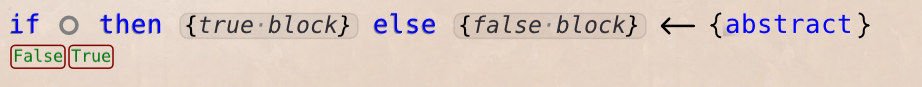}{
  The receiver between \meth{if} and \meth{then} mimics the traditional notation.
}
So the \meth{if then else} method is a method whose first argument (also receiver) is of type \type{Boolean}.
The second and third arguments after \meth{then} and \meth{else} are delayed evaluation blocks.
Note that, compared to {\smalltalk} or {\self}, the great novelty of {\omg} is that you can put a keyword {\it{} before} the receiver, but also after the last argument.
Thanks to this flexibility and simplicity when it comes to place keywords, we avoid the many criticisms of {\smalltalk} notation.
Finally, it's also possible to enclose the receiver in braces \verb|{}| for delayed evaluation.
In this way, methods with a delayed evaluation block as receiver are arranged according to their block return type context.
No more out-of-context {\it{}Block} class as in {\smalltalk} or {\self}.

Still on figure \ref{elit-boolean}, the two small red rectangles at the bottom of the method are added by {\elit} to indicate that there are two redefinitions of this method,
one in \type{False} and one in \type{True}.
By browsing these definitions, you will check that the evaluation of the \meth{then} and \meth{else} blocks respects the expected semantics.
It is particularly easy for the compiler to detect that there are only two possibilities for dynamic binding with a variable of type \type{Boolean} and
thus generate the optimal code as described in \cite{sonntag2013}.

\subsection{Properties for Operators}
All operators are library-defined.
Unlike {\smalltalk}, the {\omg} language lets you choose the priority, right-associativity or left-associativity and type commutativity for each operator.
Figure \ref{elit-andand} displays the definition of the \andand{} operator as it is defined in the \type{Boolean} prototype.
Thanks to the graphical approach, again, there's no need to invent any special syntax for defining operators.
The signature looks like any other textual method, except for the red color which is used for operators.
The right-hand side features three visual controls to select priority level, associativity, and type commutativity.
\mafigpng{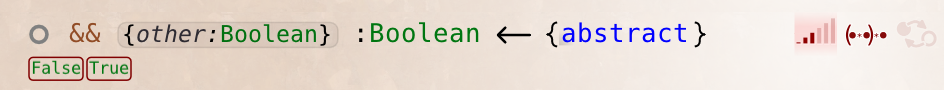}{
  Low priority level, left-associativity, and no commutativity.
}

As another example, figure \ref{elit-add} presents the definition of the {\add} operator.
The type commutativity button is enabled.
Furthermore the \local{\it{}other} argument is typed using \selftype{} to indicate that we must have exactly the same type for the argument and the receiver.
\mafigpng{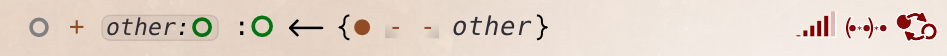}{
  High priority level, left-associativity and type commutativity.
}

Returning to the definition of the {\andand} operator in Figure \ref{elit-andand}, it's worth noting that \local{\it{}other} is enclosed in curly braces.
This indicates the presence of a code block.
Its evaluation may be delayed and, as a result, depend on the exact nature of the receiver.
In such a situation the call form must be perfectly compatible with the signature.
So, only a call of the form \local{x}\,\andand{}\,\local{\{y\}} is valid.
The aim is to fully visualize the signature choices at call site level.
This helps avoid potential errors by highlighting the possible lazy evaluation.
What's more, we've removed a special case that has no place in the elegance of a high-level visual language.

\subsection{Visualizing Clonable Slots and the Self Symbol}
As we have seen, the symbolism to indicate the position of the receiver in a signature is \selfrece{}.
This choice also makes it possible to introduce the language to beginners without having to go too quickly into the details of object-oriented programming.
At first glance, the \selfrece{} symbol might appear to be a simple routine separator
(figures \ref{elit-hello}, \ref{elit-hello2}, \ref{elit-factorial}, \ref{elit-fibonacci}).
We have also seen that within the body of a method, the \selfself{} symbol is used to designate the receiver and that its exact type is \selftype{}.
To go a step further, given that the context of the {\omg} language is prototype-based and strongly typed, there is one last important symbols we'll now introduce.
In a slot signature, symbol \selfrece{} can be replaced by symbol {\selfclone{}} to indicate the slot's behavior in the case of cloning.
Thus, the \selfclone{} symbol indicates that the corresponding slot, usually an attribute, is duplicated in the case of cloning.
\mafigpng{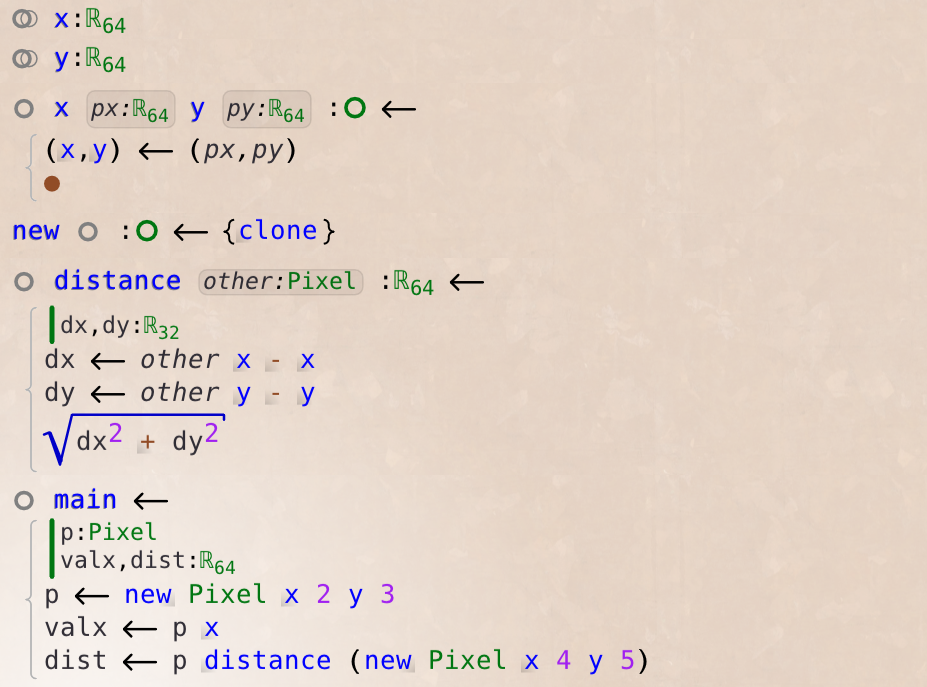}{
  The \type{Pixel} prototype example.
}
Figure \ref{elit-pixel} shows an example of defining a \type{Pixel} as being represented using two attributes \meth{x} and \meth{y}. 

Note the use of the symbol \selfclone{} in front of \meth{x} and \meth{y}, indicating the duplication of these attributes in the case of cloning.
{\omg} guarentees encapsulation by accepting attribute assignments only within the prototype that holds them (or its descendants).
In addition, when reading a slot from the outside, the code for reading an attribute or for a function call is identical.
This respects the uniform reference principle \cite{meyer-oosc}.

The ``\meth{x} \local{{\it{}px}} \meth{y} \local{{\it{}py}}'' method assigns the two attributes \meth{x} and \meth{y}.
These two attributes are assigned in a single instruction, and the method returns \selfself{} to enable a cascading call.
Still in figure \ref{elit-pixel}, next comes the definition of the \meth{new} function designed to build a new copy of \type{Pixel}.
Note here that the \meth{new} keyword precedes the \selfrece{} receiver, to get the usual creation notation.
As with the previous method, the result type is exactly that of the receiver.
The definition block only contains a clone call on the receiver, which is implicit (\selfself{} is not required).
The figure ends with the calculation of the \meth{distance} and an example of its use in
\meth{main}.

\subsection{Order of Evaluation for Graphical Operators}
As the {\omg} language is purely object-oriented, the definition of the $\Sigma$ graphical operator naturally finds its place in \type{Numeric} (figure \ref{elit-sum}).
\mafigpng{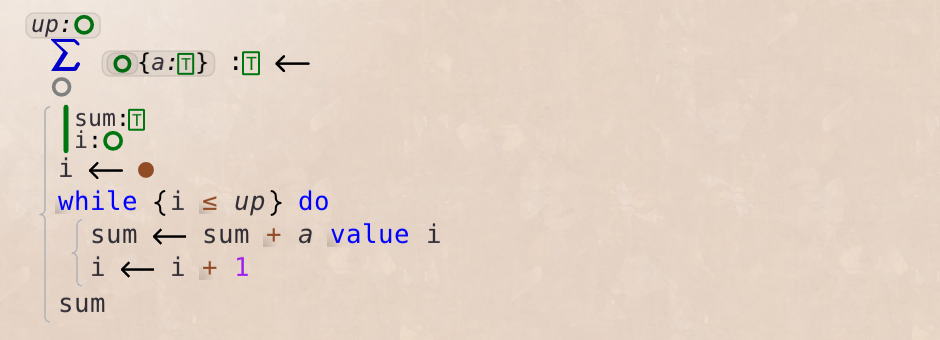}{
  Graphical $\Sigma$ operator definition.
}
The position of the receiver is below the $\Sigma$ symbol, then the \local{{\it{}up}} argument takes its place above the $\Sigma$ symbol.
The parameter \local{{\it{a}}} is a block that returns a result of a generic type parameter \pargen{T}.
Here as well, thanks to the visual effects, the generic parameters are effectively highlighted.
\mafigpng{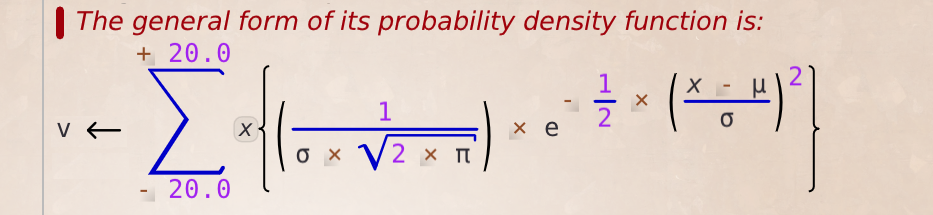}{
  Using the graphical symbol $\Sigma$ defined on Figure \ref{elit-sum}.
}
The body of the method is implemented with a loop method \meth{while do} defined in \type{Boolean}.
Figure \ref{elit-sum2} shows an example of using the $\Sigma$ operator to calculate an arithmetic mean.
The mathematical formula is as beautiful as on a school chalkboard.

Regarding the evaluation order of graphical operators, the \selfrece{} receiver is always evaluated first, just as it is the case with textual forms.
The evaluation order of the other arguments is arbitrarily determined at the same time as the definition of the vectorial shape of the operator.
Since the same graphical operator is globally defined and can appear in different contexts, the goal is to uniformize the visually perceived semantics.

\section{Visualizing software engineering aspects}
\mafigpng{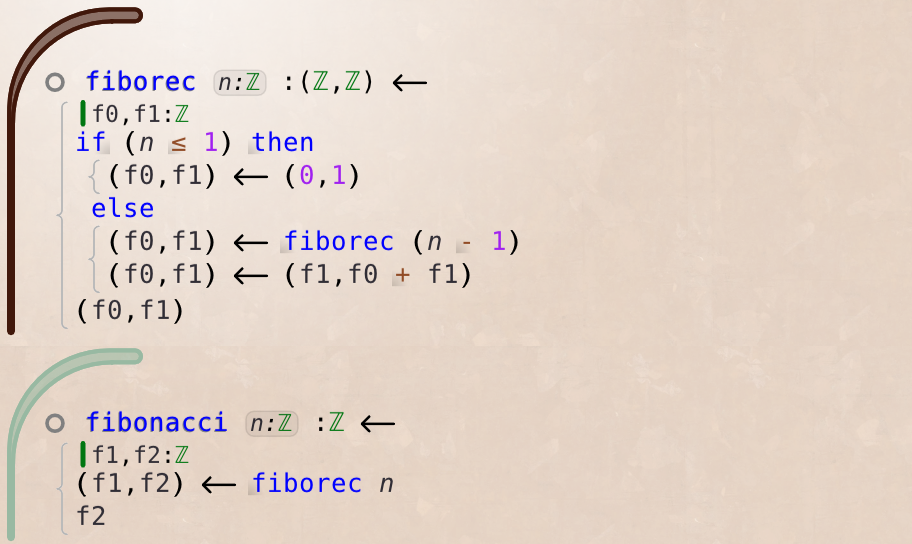}{
  Enlarged view of Figure \ref{elit-fibonacci} to show access permissions.
}
An obvious impact of the visualization layer concerns the comments placed within the code.
There is no need to choose start and end markers, such as /* and */, as is often the case.
We won't provide additional examples with comments in this article because it's easy to imagine that selecting typography,
shading, or boxing can distinguish comment sections from code sections.
Nevertheless, you can also incorporate animations, short videos, and links to URLs within the comments.
Additionally, you can refer to variables or types without the need to learn a separate sub-language, similar to Javadoc, for beautifying the comments.

\subsection{Fine Control of Slot Access Rights Also Made Visual}
The {\omg} language is designed for writing large projects and therefore requires a powerful mechanism for managing slot access permissions.
Here again, a graphic effect is used to select and visualize access permission for each slot.
Figure \ref{elit-export} repeats the example of the Fibonacci function, revealing the slot access rules with an enlarged view on the left.
The green margin on the left, enclosing the \meth{fibonacci} function, indicates that this slot is public.
Conversely, the margin surrounding the \meth{fiborec} function is dark, indicating that this slot can only be accessed by the \selfself{} receiver.
It is also possible to authorize a list of prototypes by name, or to designate a directory that factorizes access for all the prototypes it contains.
Again a visual artifact is used. No need for reserved keywords.

\subsection{Programming by Contract on the Stage}
The {\omg} language offers full support for design-by-contract programming.
Figure \ref{elit-atput} shows the abstract definition of the \meth{at put} method in \typegen{Collection}{V}.
\mafigpng{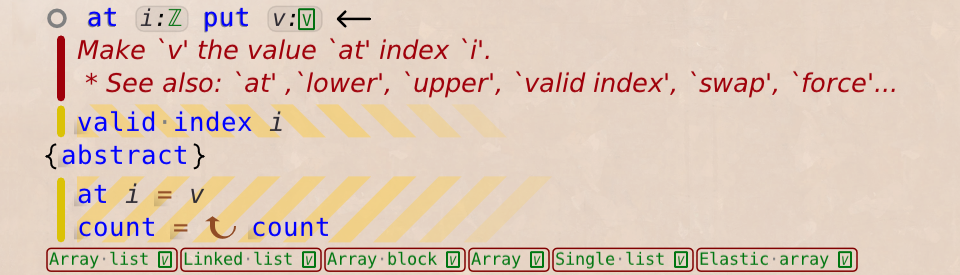}{
  Pre/postcondition visualized in the yellow hatched area.
}
Code zones corresponding to assertions are outlined in yellow hatching\footnote{Not very visible here on paper but much clearer on a real screen.}.
Here, the \meth{at put} method is used to modify a collection by replacing a \local{{\it{}v}} value at a given \local{{\it{}i}} index.
The precondition indicates that the index used must be valid within the index bounds before the call.
The postcondition indicates that, after execution, the value has been written and that the number of elements in the collection, given by the \meth{count} function, has not changed.
When assertions are enabled, code is added to check that \meth{count} gives the same result before and after the call.

The precondition and postcondition are inherited and visible in the various descendants.
This direct visibility in descendants is a major advantage over other languages. 
It is also possible, in a descendant, to add new assertions or even to ignore an inherited assertion.
Note also that {\elit} allows quick navigation to the original location where an assertion is written.
Even comments are subject to a similar inheritance mechanism.
Here again, the approach is visual and not, as in {\eiffel} \cite{meyer92a}, constrained by reserved keywords.

\section{Storage Format, Parsing, and Editing}
Each prototype is stored in a file with the same name in a format compatible with the \utf{8} encoding system.
The entire prototype is described in this single file, but we will not delve into a detailed description of its format here.
Indeed, some aspects, such as the format for comments or the format for storing all choices related to inheritance, pose no particular difficulty.
We will focus on the storage format of slot signatures, both graphical and textual, and, of course, on the format for storing method bodies.
It is in these parts that all the difficulties are concentrated.
Regarding the format of method bodies, it is important, for example, to facilitate the use of copy and paste between different parts of the code.

The goal is to reduce the size of these files as much as possible, and give the code-editing tool maximum freedom for visualization.
Our editing tool, {\elit}, itself written in {\omg}, uses a graphics library also written in {\omg}, enabling direct use of the GPU without dependence on an external library.
So all our tools are cross-platform and operating system independent.
By design, {\elit} is a demonstration of {\omg}'s ability to handle low-level programming and large applications,
while offering optimal performance without the need for C libraries to speed it up.

\subsection{File Format and Compact Source Code Encoding}
\label{file-format}
\begin{figure}[ht]
\centering
{\footnotesize\begin{tabular}{|c|c|l|}
\hline
{{\footnotesize\bf{}Symb}} & {\bf{}\utf{8} Range} & {\bf{}Description}\\
\hline
{\selfself{}} & {\footnotesize\tt{}CB99} & {{}The Receiver {\it{}aka} 'Self'}\\
{\selfrece} & {\tt{}CB91} & {{}Receiver Shared / signature}\\
{\selfclone} & {\tt{}CB90} & {{}Receiver Cloned / signature}\\
{\selftype} & {\tt{}CB9A} & {{}Exact Type of the Receiver}\\
{\includegraphics[scale=0.25]{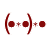}} & {\tt{}CBB1+[`0'..`9']} & {{}Left Associativity + Priority}\\
{\includegraphics[scale=0.25]{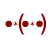}} & {\tt{}CBB2+[`0'..`9']} & {{}Right Associativity + Priority}\\
{\includegraphics[scale=0.25]{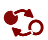}} & {\tt{}CBA0} & {{}Commutativity Switch On}\\
{}             & {\tt{}CB84/CB86} & {{}Start/Stop Superscript}\\
{}             & {\tt{}CB85/CB87} & {{}Start/Stop Subscript}\\
{\assignarrow} & {\tt{}CBBF} & {{}Assignment Arrow} \\
\hline
\end{tabular}}
\vspace{10pt}
\macaption{utf8-symbol}{The {\omg} language only reserves 64 \utf{8} codes.}
\end{figure}
Being a programming language halfway between purely textual and graphical languages, the {\omg} language requires the encoding of many meta-information:
blocks on one or more lines,
indexing or exponentiation for a specific part of a name,
multi-line or mono-line function calls, differentiation of types, local variables, and so on.
Storing information on disk must adhere to both the semantics and the choices of presentation or formatting.
In order to facilitate readability while avoiding a binary format incompatible with tools like {\git} or {\svn}, we have chosen to use tags within the source files.

An XML-type markup was quickly discarded due to the obvious file size overhead.
A lighter markup system in the style of LaTeX was attempted, but eventually, we settled on a markup using special \utf{8} characters.
The \utf{8} encoding principle allows defining $2^{21}$ characters in its normalized form, limited to 4 bytes\footnote{
Its principle could be extended up to eight bytes for a single code point, thereby increasing the count to $2^{42}$ characters.
However, for compatibility reasons, we chose to ignore this potential extension.
}.
Our idea is simply to reserve a small range of \utf{8} characters to define our tags in a minimal number of bytes.
Within the two-byte \utf{8} character set, we chose the following range, {\tt{}[0xCB80..0xCBBF]}, which represents 64 special characters reserved for {\omg}.
These are compact characters, all occupying two bytes and having no direct usage corresponding to any human language.
This lightweight approach is completely transparent to tools like {\gitsvn}, which are capable of handling standard \utf{8} encoded files.

We will not explain here the meaning of each character within this range.
It is worth noting that at present time, we utilize less than a quarter of the 64 reserved characters, thereby allowing space for future developments of {\omg}.
Figure \ref{utf8-symbol} illustrates the encoding of some basic building blocks and style tags of {\omg}.

Initially, we opted for systematic tag usage to resolve all potential ambiguities within the code of a method.
However, this approach proved to be too rigid and required challenging maintenance of tag consistency within the source file.
In fact, the validation of this markup is no longer assured when updating from an external source to the current edition, such as through {\gitsvn} updates, for example.
Furthermore, it was particularly problematic to perform copy/paste operations between different code blocks.

For all these reasons, when it comes to method bodies, we have opted for a plain representation without any tags.
This approach also facilitates the initial code typing during editing.
Consequently, with each environment loading, such as when {\elit} starts, the method bodies need to be reanalyzed.
In other words, the use of formatting tags is only present in the prototype header, for a type, or when defining a slot signature.
There is no markup within the method bodies.

It should be noted that in {\omg}, formatting an element as a subscript or superscript is not part of the namespace for identifying a slot or a type.
For example, the type \typnn{32} conflicts with a type defined as \typnnexp{32} or \typnnplain{32}.
In practice, this slight limitation is particularly useful in avoiding names that are too similar and could lead to confusion.
The same limitation is applied when it comes to method selectors.

In the definition of slot signatures, we distinguish between two types of slots: {\it{}Operator} and {\it{}Standard}.
As the name suggests, an {\it{}Operator} slot should be writable with a single symbol or unique keyword, such as \liop{$+$} or \liop{\&\&} or \liop{\tt{}||}.
An Operator slot can be unary, either pre- or post-fixed, or binary with associativity, priority, and potentially type commutativity.
A slot in the {\it{}Standard} category can have any number of arguments and is left-associative with the highest priority.
The type commutativity property applies, of course, only to slots in the Operator category.

Finally, each slot, whether in the Standard or Operator category, may matched with a graphical slot. 
If such a correspondence exists, it is not mentioned in the method bodies.
The corresponding textual representation should be used instead (detailed later in \ref{visual}).

\subsection{Parsing and Building of the Abstract Syntax Tree}
During the design of our language, we set significant requirements regarding the flexibility of writing source code.
These requirements led to significant challenges during syntax analysis. 
Aspects such as the ability to use spaces in identifiers, the absence of an explicit dot to denote dynamic binding,
and the ability to add user-defined operators through libraries made it impossible to use an LALR-type parser.

To address the inherent ambiguities in the grammar,
an upward parsing approach with dynamic programming CYK \cite{sakai1962syntax,grune2008parsing} can effectively handle context-free ambiguous grammars.
However, in the context of a strongly typed computer language, a purely context-free approach is suboptimal.
It's worth noting that the complexity of this algorithm is $O(|m|^3\cdot|G|)$, where $|m|$ is the length of the word to be analyzed, and $|G|$ is the size of the grammar.
For more precise control and reduced computational cost, we have chosen an approach that utilizes semantic knowledge to make informed rule reductions.
The parsing process occurs in two distinct phases.
The first phase, aided by the markup system, is dedicated to acquiring semantic knowledge about the entire environment (see \ref{phase1}).
It primarily focuses on prototype headers and slot signatures.
Once this knowledge is acquired for all prototypes, a second phase of parsing the plain text of slot bodies, without markup, can begin (see \ref{phase2}).

\subsubsection{First Step: Reading the Prototype Header and Slot Signatures}
\label{phase1}

\mafigpng{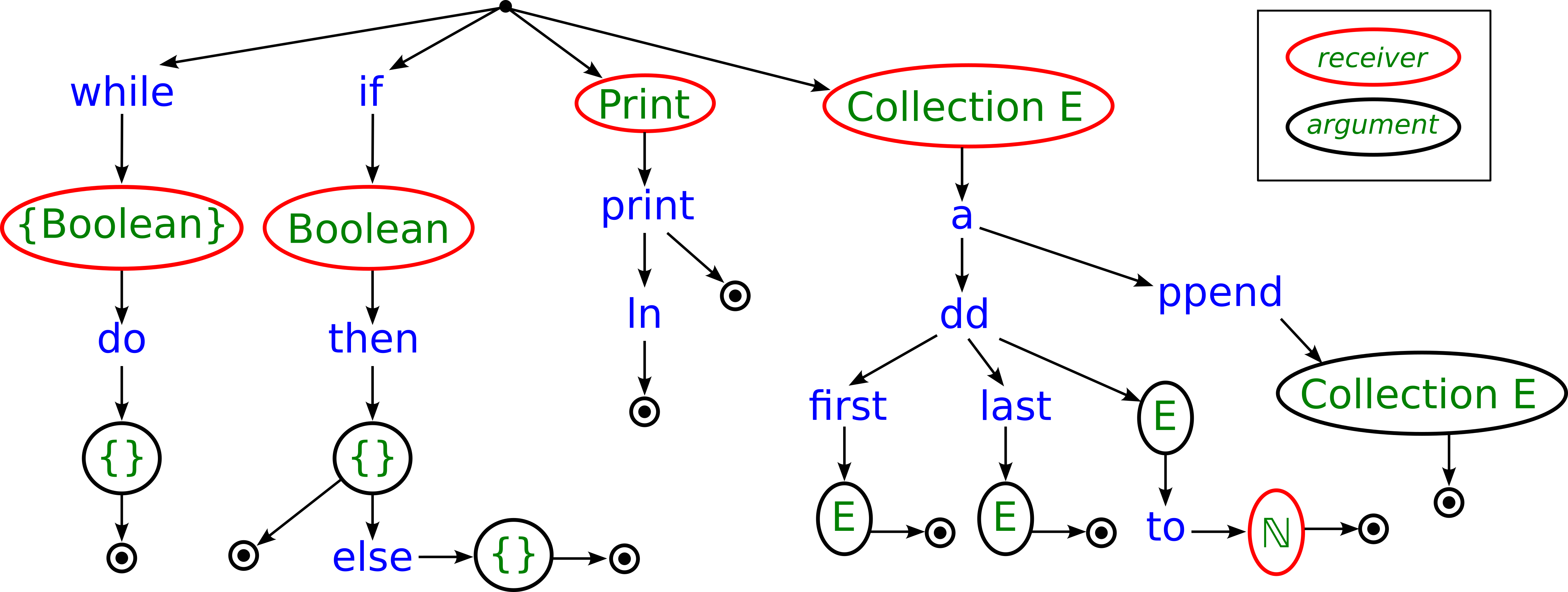}{
  Example of a syntax tree including global type information.
}  

The first phase is quite straightforward and very fast as it involves reading a pre-positioned markup system in all source files.
This allows us to locate and analyze all semantic information without ambiguity, facilitating the second parsing phase.
The parsing of the marked portion enables the rapid collection of the following information:

\paragraph{}
Complete list of prototypes with or without generality. Formatting instructions for each prototype name, subscript or superscript for each prototype name.
\paragraph{}
Directed acyclic inheritance graph, nature of links, polymorphic or not, open eyes, etc.
\paragraph{}
Signature of all Standard Slots: sequence of identifiers and formatting, presence of exponents or subscripts, receiver position, argument types, and return value types if any.  
\paragraph{}
Signature of Operator Slots: similar data collection as for Standard slots, with additional information on priority, unary pre or post-fix, or binary; priority,
right or left associativity, and commutativity type if applicable.

For Operator Slots, a global dictionary is constructed.
It will facilitate, in the second phase, the segmentation of a complex expression into multiple sub-expressions separated by operators.
Regarding Standard Slots, we build a single syntax derivation tree for all the signatures in the environment (all libraries together with the current application).
For example, for slots with identifiers such as \meth{while}, \meth{if\,then}, \meth{if\,then\,else}, \meth{add\,first}, \meth{add\,last}, \meth{add\,to}, and \meth{append},
we construct the tree shown in Figure \ref{tree-syntax}.
Thanks to this tree, we are now able to proceed to the second phase of parsing: the analysis of the plain text that makes up the slot bodies.

\subsubsection{Second Step: Parsing the Slot Bodies}
\label{phase2}
\begin{figure}[ht]%
{\footnotesize\tt\begin{tabular}{lrl}
Body    & $\rightarrow$ & \{({\it{}idf\_slot}|{\it{}idf\_local}) '\assignarrow{}'\} [\,op\,] Exp \{\,op Exp\,\}  [\,op\,]\\
Exp     & $\rightarrow$ & Atom | Msg\\
Atom    & $\rightarrow$ & \selfself{} | {\it{}number} | {\it{}string} | {\it{}external} | Type\\
        &       |       & '('\,Grp\,')' | [\,'['\,Loc\,']'\,] '\{'\,\,Grp\,'\}'\\
Grp     & $\rightarrow$ & \{\,Loc\,'$\backslash$n'\,\} [\,Body \{\,(',' | '$\backslash$n') Body\}]\\
Loc     & $\rightarrow$ & \{\,{\it{}idf\_local}\,[\,'$\colon$'\,Type\,]\,','\,\}\,{\it{}idf\_local}\,'$\colon$'\,Type\\
Type    & $\rightarrow$ & \selftype | {\it{}idf\_type}\,[\,Prm\,\{\,{\it{}idf\_type} Prm\,\}\,] \ambigfig{} \\
Prm     & $\rightarrow$ & '('\,Type\,\{\,','\,Type\,\}) | {\it{}idf\_type} \ambigfig{}\\
Msg     & $\rightarrow$ & {\it{}idf\_slot}\,[\,Arg\,\{\,{\it{}idf\_slot} Arg\,\}\,] \ambigfig{} | {\it{}idf\_local}\,[\,Arg\,] \ambigfig{}\\
Arg     & $\rightarrow$ & Atom  | {\it{}idf\_slot} \ambigfig{} | {\it{}idf\_local} \ambigfig{} \\
\end{tabular}}
\macaption{grammar}{
  The ambiguous grammar of a slot body (EBNF representation).
}
\end{figure}

The ambiguous grammar of a slot's body is presented in Figure \ref{grammar}.
The first four rules of the grammar identify assignments, declarations of local variables, blocks, constants, and tuples.
Furthermore, Operator Slots are easily recognizable thanks to the dictionary built in the previous phase.
This allows complex expressions to be divided into interleaved sub-expressions with invocations of Slot Operators.
After analyzing each sub-expression, their static type precisely identifies the invoked Operator Slots.
Building the evaluation tree, taking into account associativities and priorities, becomes possible.
However, it is important to note that a semantic error at this level of analysis is still possible.
For instance, three interleaved sub-expressions of a left-associative operator and a right-associative operator, both with the same priority, make the construction indeterministic.
In this case, the use of parentheses becomes essential to enforce the order of evaluation.

The most challenging part of the analysis lies in the terminals {\it{}idf\_local}, {\it{}idf\_slot}, and {\it{}idf\_type}.
Each use of one of these terminals corresponds to a potential ambiguity, indicated in Figure \ref{grammar} by the \ambigfig{} symbol.
The three terminals that give rise to ambiguity are:
{\it{}idf\_local} for local variables,
{\it{}idf\_slot} for Standard Slots,
and {\it{}idf\_type} for types.
It's important to note that in \omg, these three categories share the same namespace.

For parsing the content of an {\it{}idf\_*}, we use a CYK-type algorithm, utilizing the syntax tree built in the first step.
The types of parameters help reduce complexity by making cuts in the tree of possibilities.
Finally, among the different interpretations, we choose the longest recognized phrase.

In theory, it's easy to create a code example with two valid interpretations of the same size.
However, in practice, thanks to the naming convention of types starting with an uppercase letter, in contrast to lowercase for slots,
and by adhering to typing constraints, we haven't encountered any ambiguity in all the code currently written in \omg.
However, if such ambiguity were to arise, the simple addition of parentheses would be sufficient to enforce the evaluation in one direction or another, thus eliminating the ambiguity.

\subsection{Integration and Definition of Graphical Operators}
\label{visual}
In general, thanks to the knowledge of the formatting of types and slots gathered during the first phase (see \ref{phase1}),
it is the responsibility of the editor to provide a correct visualization for each use of these elements.
In the example of the type \typnn{32}, the raw written version, \typnnplain{32}, must be displayed correctly, as \typnn{32}.

All graphical slots (fraction bar, logic gate, square root, etc.) have a textual signature that identifies them.
Each graphical symbol is defined in an external file that contains not only this textual signature but also instructions for the placement of arguments
and vector drawing instructions (see Figure \ref{code-vector}).

\begin{figure}[ht]%
\centering
\begin{alltt}{\footnotesize
A0 div A1                     //{\it{} Textual corresponding form.}
v0 <- SP * 0.46
v1 <- 2
Push                          //{\it{} Draw arg #A0 and push.}
Push                          //{\it{} Draw arg #A1 and push.}
v2 <- Max(A0.Width,A1.Width)
dx <- (A0.Width - A1.Width) / 2
dy <- A1.Ascent - A0.Descent + v1 * 2
Pop (dx, dy)                  //{\it{} Pop and translate #A1.}
dx <- (v2 - A0.Width) / 2 + 5
dy <- A0.Descent-v1-v0
Pop (dx, dy)                  //{\it{} Pop and translate #A0.}
Move (Ox + 3, Oy - v0)        //{\it{} Trace the fraction line.}
Line (Ox + v2 + 7, Oy - v0)
Stroke 3
//{\it{} Return new Ascent \& new Descent.}
Ascent <- A0.Ascent - A0.Descent + v1 + v0  
Descent <- -A1.Ascent + A1.Descent - v1 + v0
Width <- v2 + 10              //{\it{} Return global Width.}
}\end{alltt}
\macaption{code-vector}{
  Vectorial operator description file for fraction.
}
\end{figure}

Without going into too much detail, we have a dictionary of graphical symbols with keys corresponding to the identifiers of a slot's signature without a type.
For example, the keys "{\bf{}A0 sqrt}" and "{\bf{}A0 div A1}" represent the symbols for the square root and fraction bar, respectively.
The visual application of the graphical symbol within a program is simply the use of a slot with a profile corresponding to the key of the symbol in question.
For example, the presence of the message in the source {\bf{}42 sqrt} will be directly interpreted graphically as follows: $\sqrt{42}$.
This solution, both simple and flexible, even allows for improving the graphical representation of old code without any modification to it.
For editing, entering {\bf{}1 div 2} is sufficient to obtain $\frac{1}{2}$.

\subsection{Global Knowledge and Static Typing for Easy Editing}
To provide the most enjoyable user experience, it is essential that the programmer can enter their code without worrying too much about formatting
(e.g. subscripts, superscripts, graphics, and indentation).
When entering source code, we avoid the use of complex keyboard commands or mouse-driven graphical interventions.
It is crucial that the programmer feels comfortable and can type their code with, at the very least, the same ease as in a purely textual language.

It is essential to emphasize that {\omg}'s choice of being a strongly statically typed language not only provides valuable parsing assistance
but is also a powerful tool for code auto-completion.
More than a constraint and somewhat paradoxically, explicit typing makes the code entry process easier and even shorter.
It should be noted that there is a slight constraint on type names, which are the only ones that can start with a capital letter.
Thus, entering an existing type name always triggers auto-completion, and a type name is most often obtained in just a few keystrokes.
Since all prototypes are preloaded by {\elit}, the absence of auto-completion indicates an error while typing the code.

For auto-completing slot calls, the syntax derivation tree built for parsing is also utilized by {\elit}.
For example, at the current time, within all the prototypes, the only slot that starts with \meth{if} corresponds to methods of \type{Boolean}.
Simply typing \meth{if} initiates auto-completion.
Another example is if you enter \meth{new}, you are presented with a list of types that have this method selector before the receiver.
Finally, by entering the initial letters of \meth{while}, auto-completion is unique and straightforward because the only possible current receiver is a block of type \type{Boolean}.
Auto-completion is semantic and is based on global knowledge of prototypes and possible signatures.

Regarding the basic elements of {\omg}, they are entered using user-redefinable shortcuts.
For example, for the assignment arrow \assignarrow, the default shortcut is '{\tt{}<-}', but someone familiar with the {\eiffel} language might opt for '{\tt{}:=}'.
Many shortcuts are obvious and almost natural, such as the shortcut '{\tt{}<=}' for '$\le$,' for example.
The default shortcut for \selfself{} is currently '{\tt{}self}'.
All shortcuts are redefinable, and they can even be associated with function keys or other specific devices.
Users simply need to make their choice according to their preferences while avoiding conflicts.

Furthermore, the syntax tree of possible signatures (similar to the one shown in Figure \ref{tree-syntax}) is annotated with probabilities of single-line or multi-line indications.
In this tree, each node corresponding to either a receiver or an argument has this information calculated based on the code encountered so far.
We use these probabilities to represent blocks during their initial creation.
For example, for the slot \meth{while} \{ \ldots \} \meth{do} \{ \ldots \}, the first block has a high probability of being on a single line,
while the second block has a high probability of being multi-line.
During auto-completion, we use this information to make the initial display choice.
Of course, the programmer has complete freedom to correct this choice.
His preference will then be stored as is and become part of the input statistics for future usage.

In the end, intuitive editing, acting directly on the visual representation of {\omg} code, is very straightforward for the user.
This is made possible in part by the minimalist nature of the {\omg} language and also by the global knowledge of the entire existing code.
What was achieved for the efficient compilation of the code presented in \cite{sonntag2013} also proves to be usable to facilitate editing.
The benefits of abandoning the archaic approach of editing a single source file and separate compilation are clearly evident in an era where
the capabilities of random-access memory easily support this comprehensive approach.
We have not discussed memory consumption regarding the consideration of all existing prototypes which are all loaded because it has remained, so far, remarkably small.

\section{Related languages or environments}
In the history of programming languages, the first language that started using visual effects for programming is most certainly {\smalltalk} \cite{smalltalk80}.
Indeed, already in 1980, all the indications concerning the inheritance are done via an interface with the mouse, using the famous browser of {\smalltalk}.
No real syntax for the classes either and only the methods are defined using a text that has to respect a fixed syntax, but otherwise very pleasant.

In {\smalltalk} the definition of control structures and conditionals is already part of the library.
Note that the operators are also redefinable, but they are all left associative with the same priority.
Huge progress, {\smalltalk} invents the keyword notation and generalizes the concept of object for all manipulated data, even the simplest.
Thus, {\omg} is in the direct line of Smalltalk's heirs, with a new, more flexible notation,
statically typed and compiled language, which swaps the classes of {\smalltalk} for the prototypes of {\self}.

In 1983, the {\prograph} language\,\cite{prograph} \cite{prograph85} is completely graphic and avoids the use of text for programs.
The underlying language is class-based and data flow driven.
The most successful implementation is the one that works on Macintosh.
The same implementation, apparently unchanged for a very long time, is still available today on MacOS X.
The greatest fault of {\prograph} is probably to have been too far ahead of his time.
Indeed, the cheap computers of that time were not powerful enough and, in any case, did not even have a mouse!

In the right line of the {\logo} language \cite{logo67}, itself non-graphical,
{\scratch}\,\cite{scratch} is probably one of the most successful visual programming environments.
{\scratch} is a simple coding language with a visual interface that allows young children to create digital stories, games and animations.
The name of the language, {\scratch}, is supposed to evoke the sound of DJs on their turntables.
The craze for {\scratch} is palpable with many users and a wide diffusion as well as multiple translations of this language.
Many videos on the web show its use and not only by children,
showing a certain interest for a less austere and more graphic programming language.

Even if {\scratch} allows for example to program the equivalent of loops and conditionals, the language remains however rather limited.
All graphical constructs are frozen and the language is essentially event driven and block assembly driven.

\mafignimp{0.90}{png}{snap}{
    The \meth{fiborec} function of figure \ref{elit-fibonacci} in {\snap}.
}
The {\snap} language\footnote{
Do not confuse {\snap}, note the use of the exclamation mark, with Snap which is a language from the 1960s
and which is dedicated to the teaching of computer science to students of humanities.
}\,\cite{snap} is largely inspired by {\scratch}.
The name is meant to evoke the noise that blocks of code make when they are assembled with each other.
{\snap} is a language that allows the definition of its own blocks, with the use of anonymous functions.
The concept of {\it{}First class sprites} corresponds to that of the prototype and {\snap} integrates all aspects of object-oriented programming.
The purely functional style can also be used with the possibility of having higher order functions.
Thus, {\snap} is a significant advance in terms of a more general-purpose visual programming language,
and not only dedicated to programming initiation for beginners or children.

Figure \ref{snap} shows the {\snap} program for the \meth{fibonacci} function of figure \ref{elit-fibonacci} following a very similar approach.
The definition is recursive but remains of linear complexity thanks to the use of a result in the form of a pair of values.
Therefore, the ability to return multiple values at once is not merely a syntactic gimmick.
Without going into details, this possibility is offered in {\omg} without ever having to allocate a container to group the different results.
The {\snap} language is very similar to {\omg} in that these two visual languages each provide full access to a real underlying programming language.
In both cases, it is no longer possible to use an ordinary text editor to edit your code.
A {\snap} block name also uses a keyword system very similar to that used for a slot name in {\omg}.
This being said, {\snap}, with its Lego-like visuals, is resolutely oriented towards the discovery of programming.
{\omg} retains a more traditional visual style that remains quite textual and integrates many aspects specific to software engineering (e.g. programming by contracts,
exportation rules, static typing, etc.).

For integrating visual aspects into the source code of programs, Jupyter Lab\cite{jupiter} is often cited as a reference.
This approach allows users to incorporate graphics, diagrams, images, and other visual elements alongside the code, thus providing an interactive and immersive experience.
However, within the scope of {\omg}, we aim to push the boundaries of this approach by introducing a new language that offers even deeper integration of visual elements.
This seamless fusion between code and graphical elements represents a significant advancement in the communication and presentation of source code,
fostering deeper exploration and a more intuitive understanding of complex concepts or calculations.

\section{Conclusion}
We have presented {\omg}, a statically typed language designed to facilitate the development of large-scale software.
In addition to ensuring safety, static typing also promotes achieving the best runtime performance \cite{sonntag2013}.
Like {\eiffel} \cite{meyer-oosc}, {\omg} integrates all programming-by-contract tools into a fully object-oriented environment.

The major innovation of the {\omg} language is to eliminate syntactic constraints (LALR grammar and parser, reserved keywords, etc.),
through the use of a visual approach.
For example, key software engineering concepts such as multiple inheritance, access rights management,
pre- and post-conditions are represented by non-fixed visual artifacts in the language.
The visualizations presented in this article (from Figure 1 to Figure 21) come from our specialized editor for {\omg}, named {\elit}, itself entirely developed in {\omg}.

In order to benefit from tools like {\git} or {\svn}, the source code of programs written in {\omg} is stored in ordinary UTF-8 text files.
Thanks to static typing and the global understanding of the entire source code, the {\omg} language parser can easily resolve ambiguities inherent in the language's flexibility.
Furthermore, global knowledge of the source code also allows for very intuitive editing of the source code, minimizing keyboard typing as well as menu usage.

Graphical operators are not limited to mathematical operators alone, and it is quite easy to add new ones.
The definition of operators is completely vectorial, and our {\elit} visualizer/editor does not use any external libraries to directly access the GPU.
All libraries used in the implementation of {\elit} are written in {\omg}.
Thus, {\elit} is the first example of a large program written with {\omg}.

%\newpage % DCDC ajustement des 2 colonnes en fin d'article
\IEEEtriggeratref{17}

\bibliographystyle{plain}
\bibliography{biblio}
%\tableofcontents
\end{document}